# Light distillation for Incremental Graph Convolution Collaborative Filtering

Xin Fan[†]  Fan Mo[‡]  Chongxian Chen[‡]  Hayato Yamana[§]

[†] [‡] Graduate School of Fundamental Science and Engineering 3-4-1 Okubo, Shinjuku-ku, Tokyo, 169-8555 Japan

[§] Faculty of Science and Engineering, Waseda University 3-4-1 Okubo, Shinjuku-ku, Tokyo, 169-8555 Japan

E-mail:  [†] fan_xin@fuji.waseda.jp,  [‡] bakubonn@toki.waseda.jp,  [‡] chenc@toki.waseda.jp

[§] yamana@yama.info.waseda.ac.jp

**Abstract**  Recommender systems presently utilize vast amounts of data and play a pivotal role in enhancing user experiences. Graph Convolution Networks (GCNs) have surfaced as highly efficient models within the realm of recommender systems due to their ability to capture extensive relational information. The continuously expanding volume of data may render the training of GCNs excessively costly. To tackle this problem, incrementally training GCNs as new data blocks come in has become a vital research direction. Knowledge distillation techniques have been explored as a general paradigm to train GCNs incrementally and alleviate the catastrophic forgetting problem that typically occurs in incremental settings. However, we argue that current methods based on knowledge distillation introduce additional parameters and have a high model complexity, which results in unrealistic training time consumption in an incremental setting and thus difficult to actually deploy in the real world. In this work, we propose a light preference-driven distillation method to distill the preference score of a user for an item directly from historical interactions, which reduces the training time consumption in the incremental setting significantly without noticeable loss in performance. The experimental result on two general datasets shows that the proposed method can save training time from 1.5x to 9.5x compared to the existing methods and improves Recall@20 by 5.41% and 10.64% from the fine-tune method.

**Keyword**  GCNs,  Collaborative Filtering,  Incremental Learning,  Knowledge Distillation

## 1. INTRODUCTION

In this age of information explosion, recommendation systems [1] [2] are playing an increasingly irreplaceable role in several real-world domains, alleviating information overload and improving the user experience. Collaborative filtering [3] is the most used approach in the field of recommender systems that attempts to capture user preferences based on explicit or implicit feedback from users on items in historical data and has been very successful in the real world [4]. In recent years, deep learning has been widely used in several research areas including natural language processing and computer vision due to its immensely powerful representation learning capabilities. Deep learning is also widely used in collaborative filtering to build rich representations of users and items to enhance the recommendation performance [5]. On top of deep collaborative filtering, graph convolutional networks (GCNs) stimulate the potential of collaborative filtering even further through their superiority of capturing higher-order interactive relationships between users and items [6][7].

Collaborative filtering based on graph neural networks has been extensively researched and much progress has been made on recommendation performance on several generic datasets [8][9]. However, in the face of huge historical interaction data, training GCNs needs to bear exponentially increasing time cost due to its high computational complexity, introduced by neighbor sampling and message passing [23][24]. The unacceptably low training efficiency prevents GCNs from being widely used in practical scenarios because it is not practical to update the model frequently to provide the up-to-date recommendations. The inability to update frequently can result in the model losing its ability to model short-term user preferences. On the other hand, updating the model using only the most recent data can lead to another problem called catastrophic forgetting [10] [11], with the model losing its ability to model users' long-term preferences.

Incremental learning has been explored in several deep learning domains to train models incrementally using the latest data while mitigating the problem of catastrophic forgetting. The two main incremental learning methods are *experience replay* and *knowledge distillation*. Methods based on experience replay [14][15] address the issue of forgetting by selecting representative data from past tasks and then training the model on new tasks using this data in combination with new data. The *knowledge distillation*-based methods [12][13] aim to preserve the model's efficacy on prior tasks by limiting alterations to its parameters. This is accomplished by incorporating task-specific regularization elements into the loss function. In this work, we focus on the methods based on *knowledge distillation*.

The key to *knowledge distillation*-based methods is how to design matching distillation terms for a specific task or a specific domain. In the field of GCNs-based collaborative filtering, several incremental learning methods based on knowledge distillation have been explored and achieved superior recommendation performance [16]. However, to achieve the desired recommended performance, these methods usually introduce additional trainable parameters or design model structures with high computational complexity leading to excessive training time consumption in the incremental setting.

In this work, we argue that current methods based on knowledge distillation over-consider distilling complex structural information as well as multiple layers of user or item representations in GCNs. The contributions of this study can be summarized as follows:

- We proposed a simple but effective light distillation method that focuses on user preferences for items in historical interactions, discarding structural information as well as representational information about users and items, thereby significantly reducing training time consumption. Specifically, we directly distilled the user's preference information by minimizing the user's preference scores for the same historical items in the previous model as well as in the current model.
- We demonstrated the effectiveness of our proposed novel distillation mothed by conducting the experiment on two commonly used datasets. Experimental results showed that our proposed distillation method can reduce training time consumption significantly and achieve competitive recommendation performance in an incremental environment.

The remainder of this paper is organized as follows: Section II introduces related studies on incremental learning. In Section III, we review the basis of GCN-based collaborative filtering and knowledge distillation. Section IV presents the proposed method. Section V presents the experimental evaluation. Finally, Section VI concludes the paper.

## 2. RELATED WORK

### 1. Incremental Learning for Deep Learning

Incremental learning is dedicated to exploring methods to periodically update the model as new data arrives. Simply updating the model with the most recent data will lead to catastrophic forgetting [10] [11], that the model loses previously acquired knowledge and shows an excessive inclination towards recent data. The central goal of incremental learning is to mitigate catastrophic forgetting. The two main lines of research on incremental learning are *experience replay* [14] [15] and *knowledge distillation* [12][13].

The *experience replay*-based methods involve sampling the most significant historical data and replaying this data during the learning of new tasks to mitigate the issue of forgetting. As a representative *experience replay*-based method. Prabhu *et al.* [12] proposed GDumb that greedily stores samples in memory as they come and trains a model from scratch using samples only in the memory at test time. The *knowledge distillation*-based methods commonly add penalty terms to the loss function to restrict the model weights from deviating excessively from their adjusted values based on historical data, thereby avoiding forgetting. Hinton *et al.* [17] first proposed transferring knowledge from a large and complex teacher model to a smaller student model without significant loss in performance. In the incremental setting, knowledge distillation is used to migrate knowledge from the previous model to the current model, thereby alleviating forgetfulness.

### 2. Incremental Learning for GCNs

Several incremental methods have been proposed to tailor the GCNs-based collaborative filtering to preserve information from different aspects. As a latest *experience replay*-based method, Fan *et al.* [6] proposed a connectivity-based sampling mechanism to enhance the graph connectivity constructed by the new date. For the *knowledge distillation*-based methods. Xu *et al.* [13] first employed knowledge distillation to preserve information from three aspects; node level, the local, graph level and the global graph level. Wang *et al.* [18] introduced a contrastive distillation mechanism for each layer of GCNs on user-item, user-user, and item-item graphs.

It can be observed that incremental learning methods based on knowledge distillation become more and more complex and cumbersome when they are tailored to collaborative filtering based GCNs. Additional trainable parameters and complex model structures improve recommendation performance to some extent, but also lead to higher training time costs. In this work, we propose a preference-driven light distillation method to align the current model with the previous model at the level of the final goal, i.e., the user's preference score for the item.

## 3. PRELIMINARIES

Before introducing the proposed methodology, we briefly review the GCN-based collaborative filtering and knowledge distillation.

*1. GCN-based Collaborative Filtering*

GCNs aim to capture the high-order collaborative signal from historical interaction data by convolution operation on the user-item bipartite graph constructed by the historical user-item interactions. For the process of forward propagation, given the initial representations $e_u$ and $e_i$ for user $u$ and item $i$, the GCNs first aggregate the information from neighbors to form the final representations $emb_u$ and $emb_i$ for user $u$ and item $i$ respectively by aggregating the information from neighbors recursively. The $emb_u$ and $emb_i$ then are used to calculate the preference score $Score_{u,i}$ of user $u$ for item $i$. Finally, the preference scores $Score_u$ of user u for the unknown items is used to form the recommendation list. In the backpropagation step, the parameters of GCNs are optimized by the Bayesian personalized ranking (BPR) loss [19], which is a pairwise loss so that a user's preference score for seen items is higher than that user's preference score for unseen items.

*2. Knowledge Distillation*

Originally, knowledge distillation aimed to transfer insights from a large, complex (teacher) model to a smaller (student) model by aligning the output logits, thereby lowering computational expenses [17]. Later studies concentrated on enhancing this transfer process and investigating its uses across different fields [12] [13]. Knowledge distillation has also been used as a mechanism to mitigate catastrophic forgetting in the field of incremental learning [10] [11]. To preserve historical learned knowledge in the previous model, the knowledge distillation mechanism minimizes the extent to which the parameters of the current model are shifted from the previous model in two successive incremental intervals by introducing additional distillation losses when training the model incrementally on new data.

## 4. METHODOLOGY

In this section, we elaborate on our proposed methodology, preference-driven distillation loss for efficiently training GCNs-based collaborative filtering models in an incremental setting. The preference-driven distillation loss mitigates catastrophic forgetting by comparing the difference between a user's preference scores for items from previous interactions in the previous model and in the current model to minimize this difference to preserve the user's historical preferences. User preference scores for items reflect the degree of user preference for items, and in GCNs, users have high preference scores for items they have seen and low preference scores for items they have not seen.

Our proposed method targets to align the user's scores for previously seen items in the previous model and the current model when the training model is in the incremental environment. Since no additional learnable parameters are introduced and the computational complexity of the model is significantly increased, it is possible to significantly reduce the training time consumption. TABLE I shows the symbols used in the study.

*A. Incremental Training*

To train the GCNs incrementally for collaborative filtering as new data arrive, we first define a set of users and items as $U$ and $I$, respectively. Second, the interactions between user $u$ and item $i$, denoted by $(u,i)$, in a dataset are sorted chronologically in an incremental setting. Besides, the interactions of user $u_i$ for an unknown item $i'$ is defined as $(u, i')$. Third, the interactions are divided into a base interval and a set of incremental intervals to simulate an incremental setting in the real world.

The observed interactions and the unobserved interactions in the t-th incremental interval is defined as $S_t^+ = \{(u,i) \mid u \in U, i \in I\}$ and $S_t^- = \{(u, i') \mid u \in U, i' \in I\}$, where $t \geq 0$. Initially, the base model is trained on $S_0$. The incremental model is trained on $S_t$ and initialized by the model trained on $S_{t-1}$, where $t \geq 1$. Fig. 1 shows the *knowledge distillation*-based incremental training procedure. The incremental model shown as $MOD_t$ distills the historical information from the $MOD_{t-1}$ by introducing the distillation loss term into the optimization objective.

TABLE I. Definition of symbols

| Symbol | Definition |
|---|---|
| $U, u$ | $U$ is the set of users in a given dataset, $u \in U$ |
| $I, i$ | $I$ is the set of items in a given dataset, $i \in I$ |
| $(u,i)$ | User–item interaction between user u and item i |
| $S_t$ | Training interactions in the t-th incremental interval, $S_t = \{(u,i,i') \mid (u,i) \in S_t^+, (u,i') \in S_t^-\}$ |
| $S_t^+$ | Oberved interactions in the t-th incremental interval, $S_t^+ = \{(u,i) \mid u \in U, i \in I\}$ |
| $S_t^-$ | Unoberved interactions in the t-th incremental interval, $S_t^- = \{(u,i') \mid u \in U, i' \in I\}$ |
| $MOD_t$ | Model trained on $S_t$ |

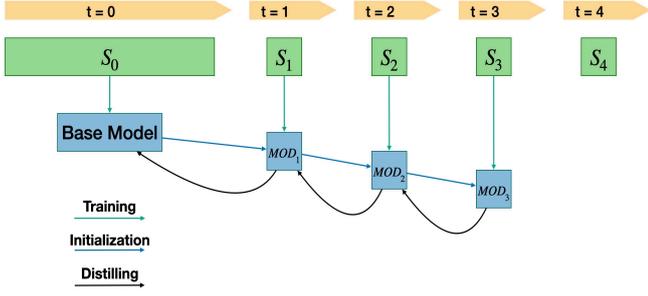

Fig. 1. The procedure of incremental learning

### B. Preference-driven Distillation

GCNs implement recommendations by learning enriched user and item representation, $emb_u$ and $emb_i$ to accurately predict users' preference scores for items. The proposed preference-driven distillation targets to distil preference information from the previous model to the current model when training GCNs in the incremental setting.

#### 1. BPR Loss

Given the final representation $emb_u$ and $emb_i$ for user $u$ and item $i$ derived by forward propagation. To optimize the parameters in a GCNs-based model, the most common choice is the pairwise Bayesian Personalized Ranking (BPR) loss [19], which enforces the prediction of an observed interaction to be scored higher than its unobserved counterparts:

$$L_{BPR} = \frac{1}{|S_t|} \sum_{(u,i,i') \in S_t} -\log \sigma(Score_{u,i}^t - Score_{u,i'}^t) \quad (1)$$

, where $S_t = \{(u,i,i') \mid (u,i) \in S_t^+, (u,i') \in S_t^-\}$ is the training data at the t-th incremental interval. The $Score_{u,i}^t$ is the dot production between $emb_u$ and $emb_i$ representing the preference score at the t-th incremental interval. The $\sigma$ is the sigmoid function. Minimizing the BPR loss makes the user's preference score for items that have been interacted with greater than the user's preference score for items that have not been interacted with.

#### 2. Distillation Loss

To preserve the user's historical preference information simply but efficiently in an incremental setting, we designed the distillation loss based on the user preference. Given the incremental model $MOD_{t-1}$ at the (t-1)-th incremental interval. We can obtain the $Score_{u,i}^{t-1}$ for each $(u,i) \in O_{t-1}^+$ that represents the user preference for the interacted item at the (t-1)-th incremental interval. The distillation loss is defined by minimizing the difference between $Score_{u,i}^{t-1}$ and $Score_{u,i}^t$ for observed interactions at the (t-1)-th incremental interval:

$$L_{KD} = \frac{1}{|S_{(t-1)}^+|} \sum_{(u,i) \in S_{(t-1)}^+} (Score_{u,i}^{t-1} - Score_{u,i}^t)^2 \quad (2)$$

, where the $|O_{t-1}^+|$ is the number of observed interactions in the (t-1)-th incremental interval. Minimizing the distillation loss allows the current model to inherit the user's preference information for interacted items from the previous model thus preserving the user's historical preference information in the incremental environment.

### C. Model Training

When training the model in an incremental setting, the BPR loss and distillation loss work together to optimize the model parameters in the process of backpropagation. The BPR loss is used to learn new user preferences on the current incremental interactions, and the distillation loss is used to inherit the historical user preferences from the previous model. The final model loss function is defined as follows:

$$L_{loss} = L_{BPR} + \lambda_1 L_{KD} + \lambda_2 |\Theta|_2 \quad (3)$$

, where $\Theta$ is the set of model parameters in $L_{BPR}$ since $L_{KD}$ introduces no additional parameters. The $\lambda_1$ and $\lambda_2$ are hyperparameters to control the strengths of distillation and L2 regularization, respectively.

---

**Algorithm I. Preference-driven Distillation**

**Input:**
Training interactions $S_t = \{(u,i,i') \mid (u,i) \in S_t^+, (u,i') \in S_t^-\}$
Oberved interactions $S_t^+ = \{(u,i) \mid u \in U, i \in I\}$
Unoberved interactions $S_t^- = \{(u,i') \mid u \in U, i' \in I\}$
Embedding of $u$ and $i$  $emb_u$, $emb_i$
**Output:** Final training loss function $L_{loss}$
**Initialize:**
Preference Score for $i$: $Score_{u,i}^t = emb_u \cdot emb_i$
Preference Score for $i'$: $Score_{u,i'}^t = emb_u \cdot emb_{i'}$
/* Definition of BPR Loss */

$$L_{BPR} = \frac{1}{|S_t|} \sum_{(u,i,i') \in S_t} -\log \sigma(Score_{u,i}^t - Score_{u,i'}^t)$$

/* Definition of Distillation Loss  */

$$L_{KD} = \frac{1}{|S_{(t-1)}^+|} \sum_{(u,i) \in S_{(t-1)}^+} (Score_{u,i}^{t-1} - Score_{u,i}^t)^2$$

/* Final Training Loss */

$$L_{loss} = L_{BPR} + \lambda_1 L_{KD} + \lambda_2 |\Theta|_2$$

The backpropagation of the parameters in the model is based on the final training loss which combines the BPR Loss and Distillation Loss

---

## 5. EXPERIMENTAL EVALUATION

### A. Datasets

To evaluate the effectiveness of the proposed method in saving the training time consumption and recommendation performance, we conducted experiments on two common publicly available real-world datasets: *Gowalla* [20] and

TABLE II. Statistical information of datasets

| Dataset | | #Users | #Items | New Users* (%) | New Items* (%) | #Interactions | Density **(%) | Timespan (month) |
|---|---|---|---|---|---|---|---|---|
| Gowalla | Base interval | 44,169 | 18,154 | | | 777,256 | 0.0969 | 19 |
| | Interval I | | | 19.7 | 1.6 | | | |
| | Interval II | | | 18.4 | 1.0 | | | |
| | Interval III | | | 16.1 | 0.7 | | | |
| | Interval IV | | | 10.4 | 0.3 | | | |
| Yelp | Base interval | 7,413 | 13,286 | | | 217,262 | 0.2206 | 24 |
| | Interval I | | | 7.1 | 7.4 | | | |
| | Interval III | | | 5.9 | 6.5 | | | |
| | Interval IV | | | 9.8 | 5.5 | | | |
| | Interval IV | | | 2.8 | 3.5 | | | |

* Percentage of cold-start users/items in each incremental interval.
** Density of each dataset is defined as (the number of interactions)/(the number of users × the number of items).

*Yelp* [21]. These datasets differ in several aspects such as the data scale, sparsity, time range, number of interactions, and numbers of new users and items. TABLE II presents the statistical details of the datasets.

- **Gowalla** [20]: Gowalla is a location-based social networking dataset collected from user check-in records, including 6,442,890 check-ins of users from February 2009 to October 2010. We filtered out duplicate interactions between users and items with fewer than 10 interaction records.
- **Yelp** [21]: Yelp is an all-purpose dataset provided by the Yelp business and includes 6,990,280 user check-in records from February 2005 to January 2022. We utilized data from the period 2017-2018 and applied the same filtering method as that used for the Gowalla dataset.

B. *Experimental Setting*

Following the temporal order, we divided each dataset into a base interval, containing 60% of the entire dataset, and four incremental intervals, each containing 10% of the entire dataset. The process of incremental training is described in Figure I. The base model is trained on $S_0$ at t=0 and the incremental model is trained on $S_t$ sequentially. The model trained on $S_{t+1}$ as training data is validated and tested on $S_{t+1}$ as validation and test data.

**Base Model:** The proposed incremental learning method is built on a base GCNs-based model. In this work, we choose LightGCN [22] as our base model, which is widely used in the field of collaborative filtering-based recommender systems. LightGCN removes the nonlinear activation functions and learnable parameters for linear transformation of traditional GCNs, which significantly simplifies the training burden of GCNs and achieves the desired recommended performance.

**Baselines:** To compare the effectiveness of our proposed method in reducing training time and its recommended performance. We compare our method with the following knowledge distillation-based baselines in terms of both training time consumption and accuracy:

- *Fine Tuning*: The model trained in $S_t$ initializes the trainable parameters by the parameters of the model has been trained in $S_{t-1}$ with no additional distillation loss. The most naive method which is often used as a benchmark to compare the effectiveness of other incremental learning methods.
- *Graph SAIL* [13]: Information at the node, local and global levels in the GCN is distilled through three customized distillation losses, respectively. Among them, clustering of user and item representations is required for distilling information at the global level.
- *LWC* [18]: The introduction of a comparative distillation mechanism combines BPR losses with distillation losses. The item-item and user-user graphs are introduced on top of the traditional user-item bi-directional graphs and the information is distilled layer by layer.
- *Connectivity Sampling* [6]: The connectivity of graph structure is considered in the incremental setting. The incremental learning method based on experience replay by sampling representative historical

TABLE IV. OVERALL EXPERIMENTAL RESULTS ON THE TOP 20 RECOMMENDED ITEMS

| DataSet | Methods | Interval I | Interval II | Interval III | AVG R@20 | AVG Improvement from Fine Tune (%) | AVG Runtime (s) |
|---|---|---|---|---|---|---|---|
| Gowalla | Fine Tune | 0.1065 | 0.1167 | 0.1206 | 0.1146 | 0 | **2,428** |
| | Graph SAIL [13] | 0.1089 | 0.1181 | 0.1244 | 0.1171 | 2.18% | 14,345 |
| | LWC [18] | 0.1156 | 0.1218 | 0.1281 | 0.1218 | 6.28% | 40,185 |
| | Connectivity Sampling [6] | 0.1181 | 0.1245 | 0.1309 | 0.1245 | **8.63%** | 10,471 |
| | **Proposed** | 0.1149 | 0.1191 | 0.1283 | 0.1208 | 5.41% | 4,503 |
| Yelp | Fine tune | 0.0559 | 0.0589 | 0.0576 | 0.0575 | 0 | **952** |
| | Graph SAIL | 0.0603 | 0.0664 | 0.0612 | 0.0626 | 8.87% | 7,997 |
| | LWC [6] | 0.0627 | 0.0661 | 0.0632 | 0.0640 | 11.31% | 10,853 |
| | Connectivity Sampling [6] | 0.0642 | 0.0675 | 0.0635 | 0.0651 | **13.21%** | 3,046 |
| | **Proposed** | 0.0611 | 0.0672 | 0.0624 | 0.0636 | 10.61% | 2,047 |

interactions to enhance the connectivity of the graph structure of the current model.

**Hyperparameters:** The tuning process followed a greedy strategy. The learning rate was tuned with *fine tuning* in the range of {0.005, 0.001, 0.0005, 0.0001} and then fixed to 0.0005 for the model trained on $S_0$ and 0.0001 for the model trained on $S_t$ for other baselines. The weights of the different loss terms for Graph Sail [13] were tuned by a grid search in the range of {0.1,1,100} for $\lambda_{self}$ and {1e4,1e5,1e6,1e7} for both $\lambda_{local}$ and $\lambda_{global}$. The hyperparameter $\tau$ for *LWC* [18] was searched in the range of {0.1,0.2,0.5,1}. The hyperparameters $\lambda_1$ for our proposed method were tuned in the range of {0.01,0.1,1,10,100}. TABLE III presents the values of the final hyperparameters.

*C. Comparison*

The proposed distillation method was compared with the baselines in terms of both recommendation performance and training time. Recall on the top 20 recommendation items was used as the metric to evaluate the recommendation performance. For the evaluation of training time, we calculate the time in seconds it takes for the model to reach the best Recall. TABLES IV shows the experimental results for the dataset Gowalla and Yelp, respectively. From the experimental results, we can conclude the following:

**Recommendation Performance:** The proposed method can comprehensively outperform *Graph SAIL* [13] in each incremental interval evaluated by Recall@20 on the Gowalla dataset. Compared to fine tune method, our proposed method achieved an average improvement of 5.41% over three incremental intervals, which fully demonstrated the effectiveness of the proposed method despite its simplicity. Although there was still a gap compared to *LWC* [18] and *Connectivity Sampling* [6], this is acceptable in the face of the expensive cost of training time at *LWC* which employs a complex distillation mechanism to distill information from each layer of multiple graph structures.

**Training Time:** The training time mainly depends on the number of parameters of the model, the time complexity of the model, and the number of epochs required for the model to reach convergence. The experimental evaluation for training time was performed on an NVIDIA A100 Tensor Core GPU by calculating the product of the number of epochs required to achieve the best recall@20 and the time required for each epoch. The results on the two datasets show that our proposed method can significantly reduce the time consumed for model training from 1.5x to 9.5x compared to these baselines.

*D. Ablation Study*

To explore the effect of the degree of distillation on the recommended performance, we conducted experiments on the hyperparameter $\lambda_1$ in the range of {0.01,0.1,1,10,100} on both *Gowalla* and *Yelp* datasets. TABLE. V shows the experimental results for the average result of Recall@20 for three incremental intervals. Because the level of distillation does not significantly affect training time, we do not discuss training efficiency here. We can observe that the effect of hyperparameter $\lambda_1$ on recommendation performance can reach a maximum of 7% on the Gowalla dataset and 30% on Yelp dataset, which indicates that fine-tuning the hyperparameter $\lambda_1$ is of practical importance.

TABLE III. Hyperparameters

| | $\tau$ | $\lambda_{self}$ | $\lambda_{local}$ | $\lambda_{global}$ | $\lambda_1$ | $\lambda_2$ |
|---|---|---|---|---|---|---|
| Gowalla | 1 | 100 | 1e5 | 1e7 | 1 | 1e-3 |
| Yelp | 0.5 | 0.1 | 1e3 | 1e2 | 0.01 | 1e-3 |

TABLE III. Study for Hyperparameters $\lambda_1$ on the Recall@20

| | *0.01* | *0.1* | *1* | *10* | *100* |
|---|---|---|---|---|---|
| Gowalla | 0.1145 | 0.1128 | 0.1149 | 0.1145 | 0.1074 |
| Yelp | 0.0603 | 0.0563 | 0.0553 | 0.0494 | 0.0462 |

# 6. CONCLUSION

In this work, we propose a novel incremental learning method based on knowledge distillation that can efficiently train collaborative filtering models based on GCNs and achieve competitive recommendation performance. To mitigate the catastrophic forgetting problem in incremental environments, we propose methods to distill the user's historical preference information from the previous model to the current model. Experimental results on two publicly available datasets show that the proposed method can reduce the training time from 1.5x to 9.5x compared to the existing methods. With such time consumption, the proposed method can achieve competitive recommendation performance, improving Recall@20 by 5.41% and 10.61 respectively on two datasets.